\documentclass{webofc}
\usepackage[varg]{txfonts}   
\usepackage[labelfont=bf]{caption}
\usepackage{subcaption}
\usepackage{makecell}
\usepackage{natbib}
\begin{document}
\title{Convolutional LSTM models to estimate network traffic}
%
%

\author{\firstname{Joanna} \lastname{Waczyńska}\inst{1,2}\fnsep\thanks{\label{main}main authors} \and
    \firstname{Edoardo}
    \lastname{Martelli}\inst{2}\fnsep\textsuperscript{\ref{main}} \and\firstname{Sofia} \lastname{Vallecorsa} \inst{2}\fnsep\textsuperscript{\ref{main}}  \and
    \firstname{Edward} \lastname{Karavakis}\inst{2} \and
    \firstname{Tony } \lastname{Cass}\inst{2}
}

\institute{\footnotesize Wrocław University of Science and Technology (Wroclaw, Poland) -- email: \texttt{joanna.waczynska@gmail.com}
\and
CERN (Geneva, Switzerland), IT department -- email: \texttt{firstname.lastname@cern.ch} 
}

\abstract{%
 Network utilisation efficiency can, at least in principle, often be improved by dynamically re-configuring routing policies to better distribute on-going large data transfers. Unfortunately, the information necessary to decide on an appropriate reconfiguration---details of on-going and upcoming data transfers such as their source and destination and, most importantly, their volume and duration---is usually lacking. Fortunately, the increased use of scheduled transfer services, such as FTS, makes it possible to collect the necessary information. However, the mere detection and characterisation of larger transfers is not sufficient to predict with confidence the likelihood  a network link will become overloaded.  In this paper we present the use of LSTM-based models (CNN-LSTM and Conv-LSTM) to effectively estimate future network traffic and so provide a solid basis for formulating a sensible network configuration plan.
}
\maketitle
\section{Introduction}
\label{intro}
The evolution of data engineering has increased the frequency of use of applications and data transmission services. Most transfer services assume that the network is always ready for their transfers and this behaviour has increased the number of clashing transmissions that compete for the scarce available bandwidth.

\begin{figure}[h]
\centering
\begin{minipage}{.5\textwidth}
  \centering
  \includegraphics[width=0.9\linewidth]{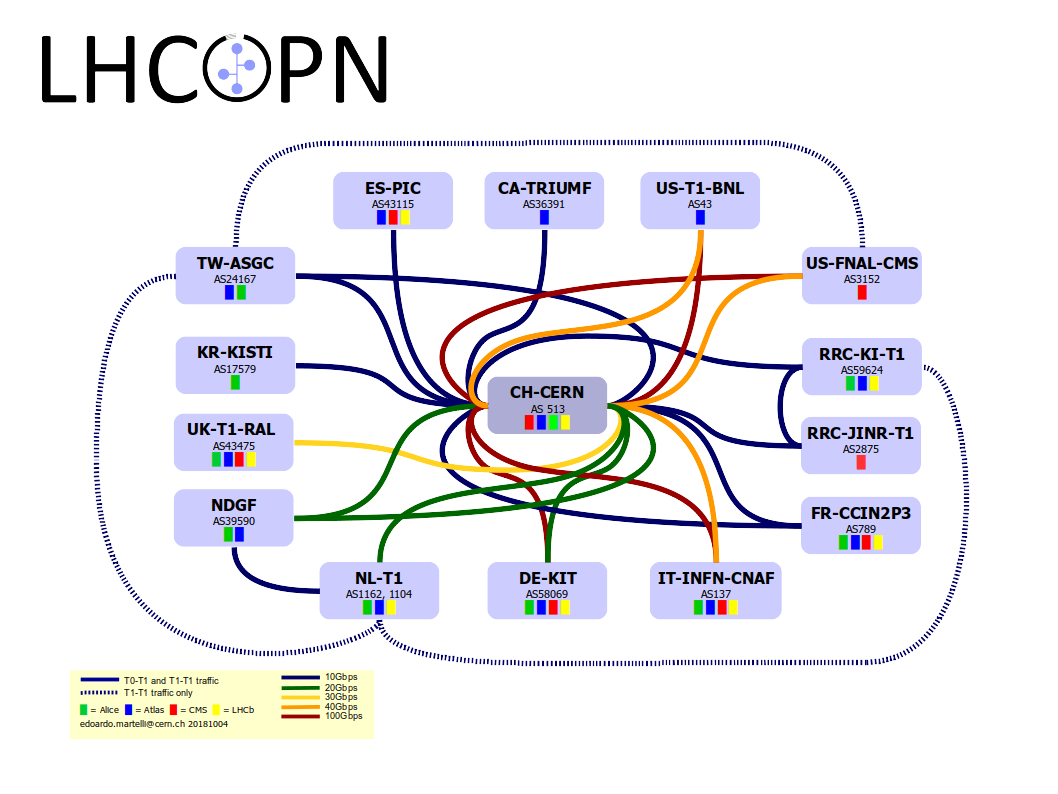}   
    \caption{LHCOPN topology} 
    \label{fig:lhcopnmap} 
\end{minipage}%
\begin{minipage}{.5\textwidth}
  \centering
\includegraphics[width=0.9\linewidth]{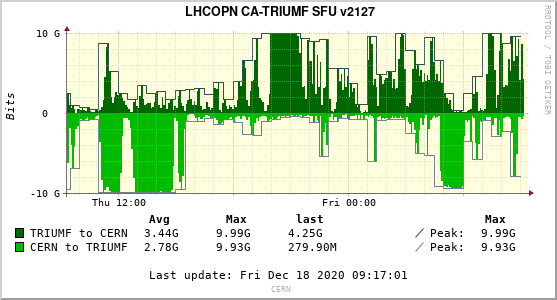}
\caption{Network traffic observed on the LHCOPN path between CERN and TRIUMF. Link saturation occurs in both directions.} 
\label{fig: exapmpleSaturation} 
\end{minipage}
\end{figure}

This work analyzes data provided by FTS \cite{FTS}, the most important source of traffic on the LHCOPN (Large Hadron Collider Optical Private Network). The LHCOPN topology is presented in figure \ref{fig:lhcopnmap}.  In FTS, data transfers are reported relatively continuously, without major delays or interruptions. Our analysis is based on a time series with a sampling frequency of two minutes, chosen taking into account the frequency of  information reporting between two storage elements by FTS. Figure \ref{fig: exapmpleSaturation} shows the network traffic observed between the CERN and TRIUMF WLCG sites, one of the data transfer paths discussed in this paper. It is not uncommon to observe link saturation during data transfer, when network traffic occupies almost all the available bandwidth. The possibility to dynamically configure the network to add bandwidth and load-balance traffic could reduce occurrences of link saturation and so improve transfer performance and prevent recoverable (server or networking) errors~\cite{myNoPublic}. Detecting and understanding transfer behavior, through Deep Learning, is a key part of this endeavour.  Machine Learning and Deep Learning algorithms are being investigated in different fields of industry and science to improve the optimisation of complex dynamical systems. The key benefit of using these methods is the automation of the model life cycle steps \cite{pipe}. In this context, it becomes essential to design a Deep Learning model for traffic estimation \cite{8717920}, ensuring a flexible and adaptive learning process.

FTS makes it possible to perform transfers between any two random endpoints reliably and efficiently~\cite{FTSdoc}. Given the behaviour of the FTS Optimizer, transfers on an empty or lightly-loaded network path show self-similarity. An exception is the occurence of sudden throughput drops caused by unexpected errors. Fortunately most of these are quickly rectified by FTS so that when observing the network traffic we do not see a long-term problem. Therefore, a historical time series analysis is a natural choice for making traffic predictions~\cite{gas} -- segment analysis can bring  greater accuracy in the forecast of network traffic, since, by examining the segments, one is able to detect incorrectness or minimize the number of reading errors.  Real Tier1 data (i.e network traces, sequences and time windows) can be used to estimate the current network traffic and predict future behaviour~\cite{8717920}.

Since our work is aimed at detecting the saturation of the link and the effective end of the problem, analysing the multiple features affecting traffic and obtaining reliable time forecasts is essential.
While convolutional layers are used to extract correlations across multiple features,  Recurrent Neural Network (RNN), and especially long short-term memory (LSTM), can interpret network data as time sequences  \cite{convlstmeq, lsdtmTime} and, indeed,  LSTMs have been shown to outperform traditional RNNs in numerous temporal processing tasks~\cite{lstm}.
Considering the importance of the (short) history data, a combination of CNN and  LSTM layers becomes the natural architectural choice.
 This work proposes a Deep Learning based method to estimate network utilization (network traffic) while comparing four different models: CNN, LSTM, CNN-LSTM~\cite{CNNLSTM,livieris2020cnn}, and Conv-LSTM~\cite{zheng2020hybrid}. 
 
 The paper is organised as follows: the next section presents  the data sets, a description of the data pre-processing step and the DL models follows, together with an analysis of the results leading to the choice of the best model.

\section{The input data sets}
\label{sec-inputdataset}
This study focuses on two LHCOPN links, between CERN and TRIUMF and between CERN and PIC, both with a capacity of 10Gbps (figure \ref{fig:lhcopnmap}).
Our data sets are generated by the  NOTED software \cite{myNoPublic} which detects and aggregates all FTS transfers on a link under study.
It should be noted that FTS is not the only source of network traffic; in order to reduce external influences, therefore, we take care to analyse a time period when most of the traffic is generated by FTS.

For short duration transfers, the most significant transfer metric is the  FTS throughput (TH), which is adjusted by the FTS optimizer \cite{FTSdoc}. During  long transfers, throughput should asymptotically approach the link bandwidth, but transmission problems can cause drops that do not correlate linearly with the observed traffic. We take care to include such cases in our analysis. 

\begin{figure}[h] 
\centering
    \includegraphics[width=0.5\linewidth]{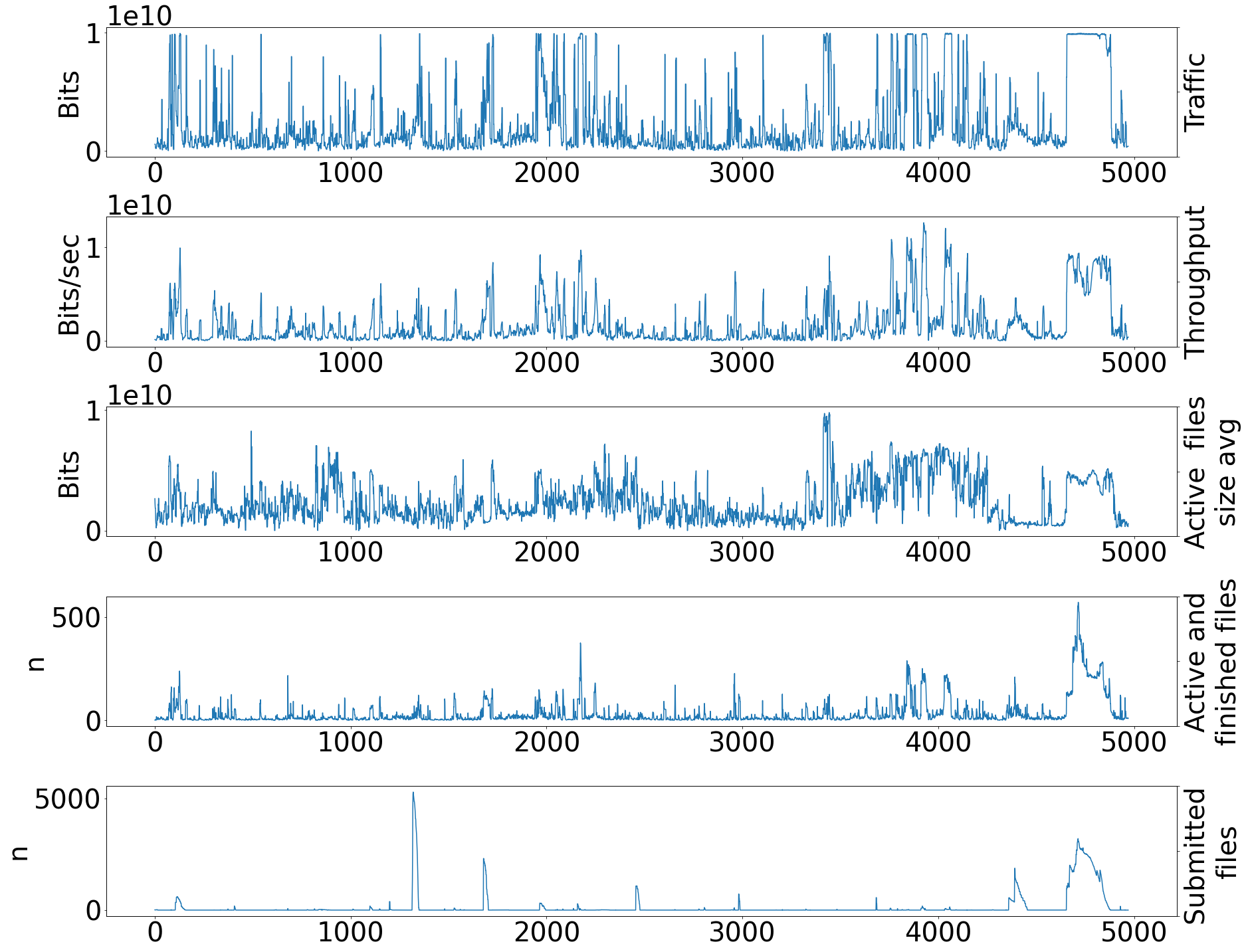}     
\caption{Training data set collected on the TRIUMF to CERN path, including: throughput, average active files size, sum of active and finished files, submitted files. } 
    \label{fig: processed} 
\end{figure}

After the transfer is loaded, FTS groups all files into similar sized jobs. Thanks to this, during the transfer, we expect the average size of active files to be constant. Such a relationship is shown in figure \ref{fig: processed}. Given that only the total number of active and finished files is available, we interpret this value as an estimator of the number of active files (AF). 

Once a link is congested, the submitted files are queued. Thus, estimating the end of a transfer corresponds to predicting the time to empty the FTS queue  plus the time necessary to transfer the remaining active files \cite{myNoPublic}. Figure \ref{fig: processed} shows the network traffic and the corresponding information on aggregated transfers. 

\begin{figure}[h]
\centering
\begin{subfigure}{.47\textwidth}
  \centering
  \includegraphics[width=\linewidth]{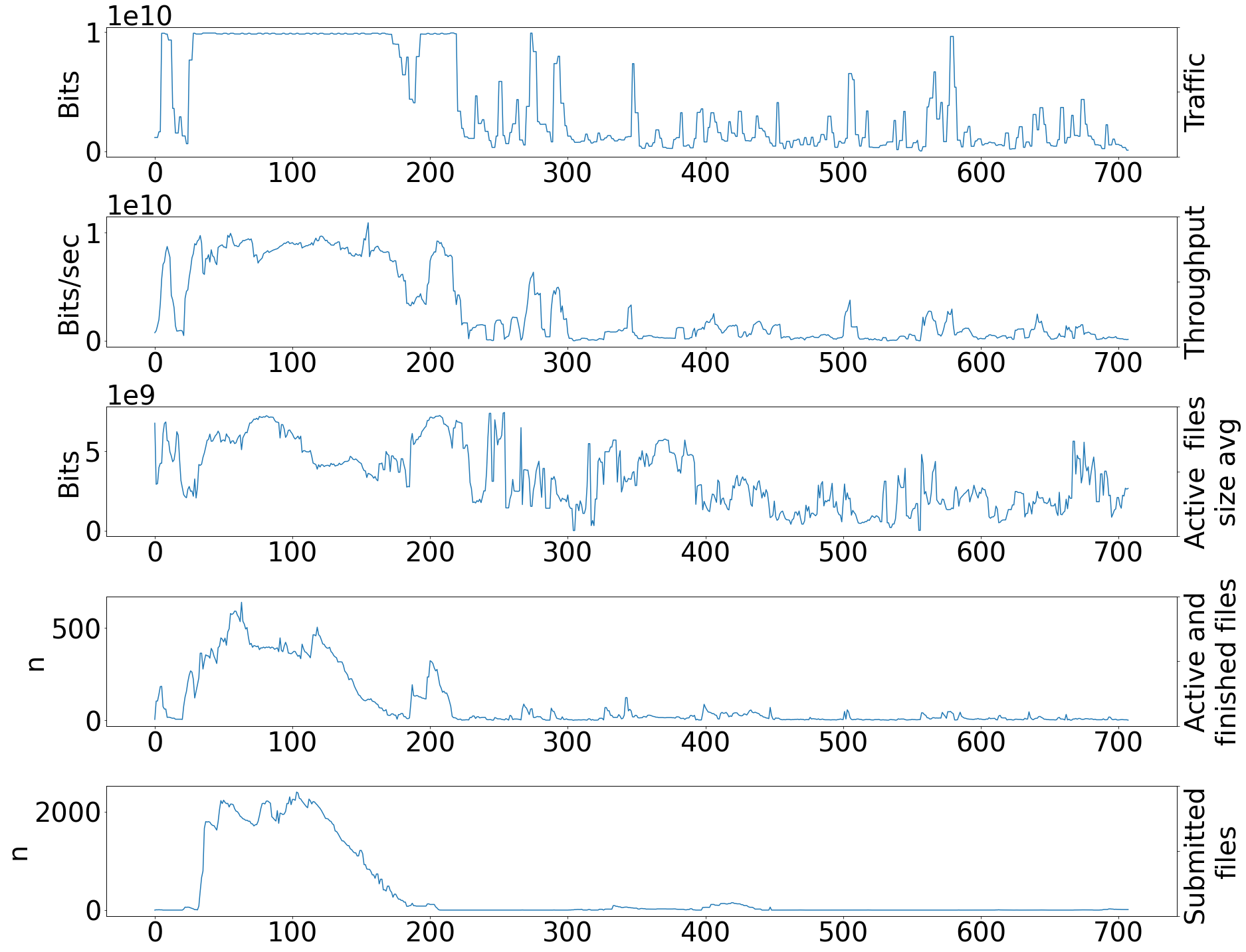}
  \caption{Transfers from TRIUMF to Tier0/Tier1 sites via CERN.}
  \label{fig:testsub1}
\end{subfigure}%
\hspace{0.5cm}
\begin{subfigure}{.47\textwidth}
  \centering
  \includegraphics[width=\linewidth]{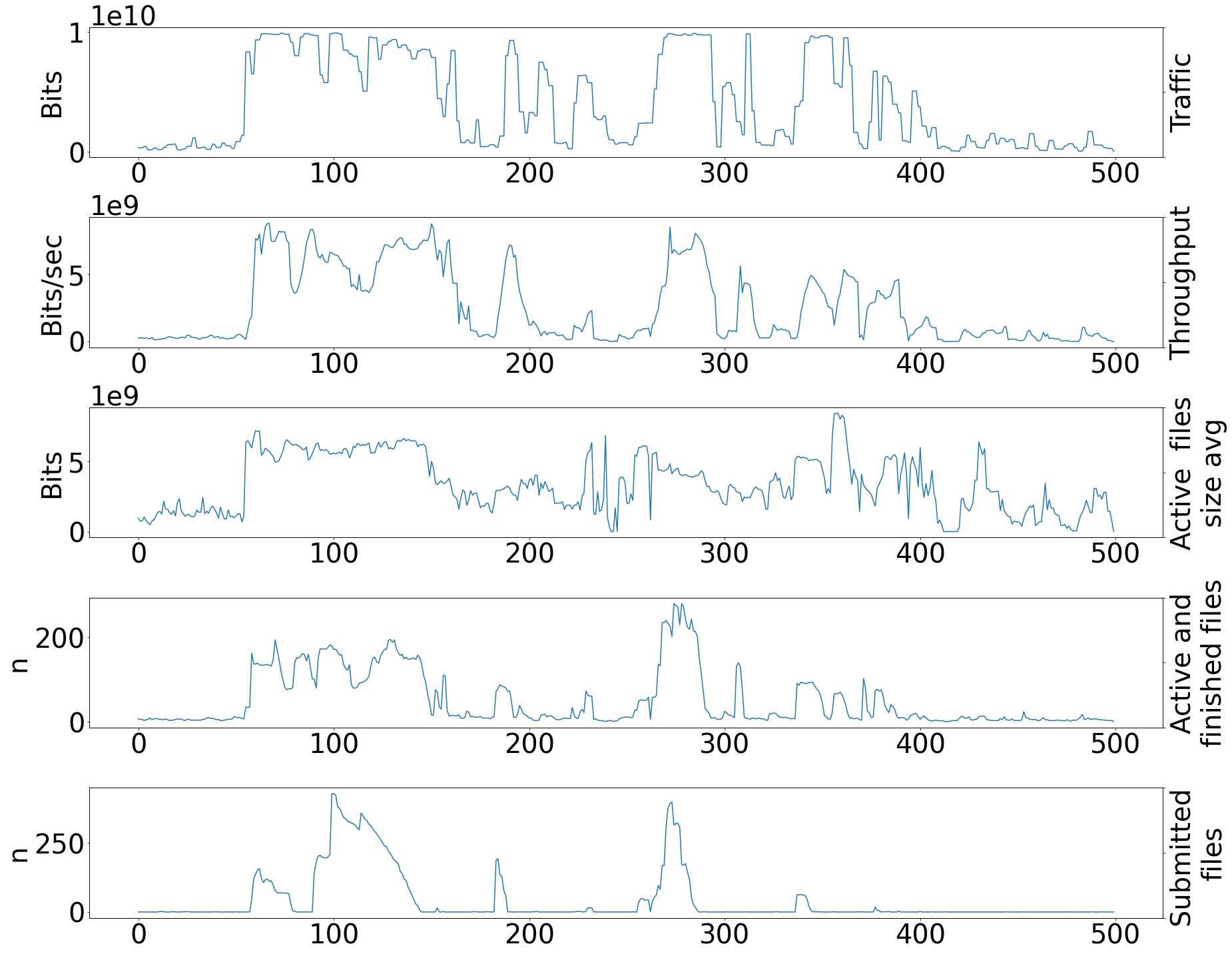}
  \caption{Transfers from Tier0/Tier1 sites to TRIUMF via CERN.}
  \label{fig:testsub2}
\end{subfigure}
\caption{Test data sets to/from TRIUMF.}
\label{fig:test}
\end{figure}

The CNN, LSTM and CNN-LSTM models are trained using the following set of parameters (presented in figures~\ref{fig: processed},~\ref{fig:testsub1} and \ref{fig:testsub2}): the throughput (TH), the number of active files for a given time step (AF), i.e. the number of files being transported by FTS; the queue length (QL), or number of submitted files (SF) and, finally, the average active files  size (AFS), that were processed since the last FTS report. For the Conv-LSTM model we consider additional pre-processing steps and calculate the throughput exponential moving average over the last 15 minutes, and estimated size values. The AF to AFS multiplication and the QL to AFS multiplication are approximations of the active files size and the queue size respectively.

We analyse periods of both normal load and saturated utilisation of the links. 
A normal period is characterized by sudden increases of the link utilisation and equally rapid decreases, with sometimes short-lasting saturation. During saturation the link is  used at its maximum capacity,  concurrent transfers are slowed down and any additional transfers will degrade overall performance. Saturation can be due to single, large  transfers (figure \ref{fig:testsub2}, equivalent to the 1st~queue), or to the accumulation of many concurrent small transfers (figure \ref{fig:testsub1}).

The training data set uses transfers originating from TRIUMF that are directed to CERN or that pass via CERN to reach another Tier1 site~\cite{myNoPublic} connected to the LHCOPN. 

To study the generalisation capability of our models we choose test and validation data sets collected in different configurations.
The first test data set (figure \ref{fig:testsub1}) is collected in exactly the same way as the training data, but during a different time period. These data describe aggregated information regarding the outgoing transfers from TRIUMF to CERN(Tier0) and to T\mbox{ier1} sites reached by transiting CERN. We use performance on this data set to determine which of our models is the best.
Next, we analyse transfers along the same path but in the opposite direction, i.e. from Tier1 / CERN to TRIUMF (figure  \ref{fig:testsub2}) and   transfers  from PIC to CERN and Tier1s and from CERN and Tier1s to PIC, during the same time period  as  the TRIUMF data set.

The traffic in figure \ref{fig:testsub1} is caused by many transfers, as shown by the queue size ("submitted files" in the figure) which initially decreases,  then increases as new transfers/files are injected into FTS. Note the gradual decrease in the number of active files, despite the existence of a queue. This data set shows perfectly the relationship between throughput and network traffic: when the FTS throughput decreases so does the network traffic and  the saturation is reduced.
The second data set, Tier0/Tier1 to TRIUMF  (figure \ref{fig:testsub2}), exhibits a different transfer pattern, resulting in four different queues (shown by the submitted file panel in figure \ref{fig:testsub2}).  Four large transfers are present which, unlike the first data set, do not overlap (except for the 2nd queue). It should be noted that the observation of the 3rd queue does not correspond to an increase in the number of active files, despite the increased FTS throughput. The second and fourth queues have similar sizes, and correspond to similar throughput values, but the corresponding transfers end much faster. This is due to the occurrence of additional bulk transfers.

\section{The data pre-processing step and DNN architectures}\label{mse}

The goal of the NOTED project is to study the relationship between FTS transfers and network traffic in order to trigger actions of an SDN controller to alleviate saturation \cite{myNoPublic,inproceedings}. Therefore we focus on traffic estimation at the time of link overload, defining two main tasks: forecasting the immediate and the short-term evolution of network traffic.

\begin{figure}[h]
\centering
\begin{minipage}{.45\textwidth}
  \centering
  \includegraphics[width=0.75\linewidth]{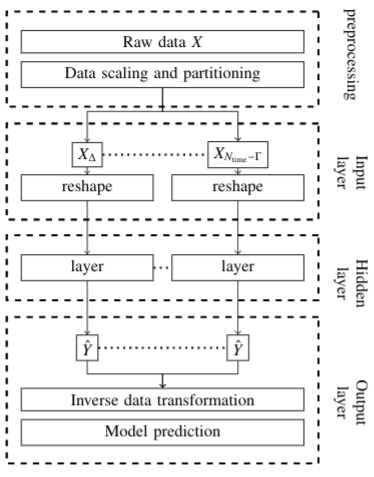}
  \captionof{figure}{\vspace{-0.7cm }General processing model.}
  \label{fig:ntframework}
\end{minipage}%
\hspace{0.5cm}
\begin{minipage}{.45\textwidth}
  \centering
  \includegraphics[width=0.8\linewidth]{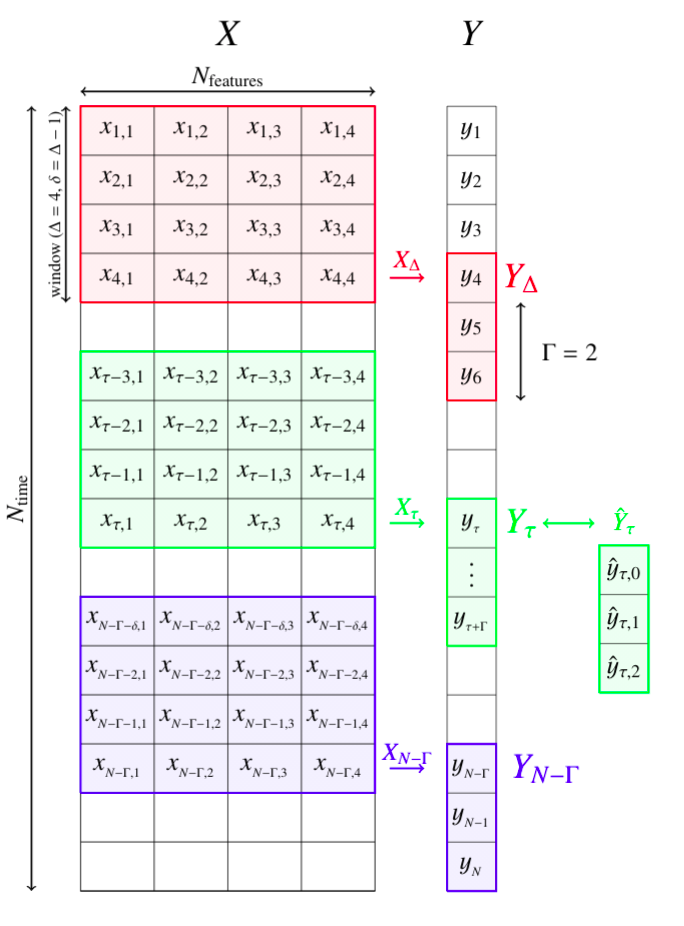}
  \captionof{figure}{Schema input data set $X$, and dependent variables $Y$. }
  \label{fig:equationSchama}
\end{minipage}
\end{figure}

Section \ref{sec-inputdataset} introduced the features we use for training our models. Considering the cross-correlations between them and their temporal auto-correlation, we choose models built as a combination of convolutions and LSTM units. We employ a supervised training approach and we define the training data set $(X,Y)$, as in figure \ref{fig:equationSchama}.
We consider  $N_{\text{time}}$ steps, a time window  $\Delta$ and $\Gamma$ forecasting steps. We choose a time window $\Delta \in \{ 4,10\}$ taking into account the dynamics of traffic measurements. Given that measurement errors can generate artificial drops generally shorter than 10 mins, we assume that an 8 mins time window ($\Delta = 4$) is a reasonable choice. However, FTS can automatically identify and recover from  unexpected traffic drops and as we have observed that such a recovery can take up to 20 minutes,  we choose $\Delta = 10$ (20 minutes). The $\Gamma$ value depends on the network controller needs and it is discussed in section \ref{sec-analis}. The raw input data is represented by $X_{\text{raw}}=\{x_{i,j}\}$ where $i=1\dots N_{\text{time}}$ and $j=1 \dots N_{\text{features}}$ and the time window is interpreted as the input segment to the models. The pre-processing procedure is aimed at  grouping the data features according to time windows and its output is a set of matrices 
${X=[X_{\Delta}, \dots, X_{\tau} , \dots, X_{N_{\text{time}}-\Gamma }]}$, each with dimensions $N_{\text{features}} \times \Delta$:

\begin{equation}\label{eqn:outputfortau}
X_{\tau} = \begin{bmatrix} 
    x_{\tau-
    \Delta+1,1} & x_{\tau-\Delta+1,2} & \dots \\
    \vdots & \ddots & \\
    x_{\tau,1} &        & x_{\tau, N_{\text{features}}}
    \end{bmatrix}
\end{equation}
The true traffic measurements form a vector of size $\Gamma$, $Y_\tau = [y_\tau,\dots,y_{\tau +\Gamma} ]$,  associated to each $X_\tau$ matrix. The estimated values at time $\tau$ is $\hat{Y}_\tau = [\hat{y}_\tau,\dots,\hat{y}_{\tau +\Gamma} ]$.

The general scheme of our networks is presented in figure \ref{fig:ntframework}, its main components include: the pre-processing step described above; a input layer containing the input segments ${X=[X_{\Delta}, \dots, X_{\tau} , \dots, X_{N_{\text{time}}-\Gamma }]}$ and an additional reshaping operation in the case of Conv-LSTM and  the hidden  layers representing the different core architectures: LSTM units in the pure LSTM model and one CNN layer and one LSTM layer for the CNN-LSTM. The  Conv-LSTM model also includes a flattening layer to adjust the output dimensions. Finally an output layer predicts the variables  defined by $\hat{Y}_{tau} = [\hat{y}_\tau,\dots,\hat{y}_{\tau +\Gamma} ] = f\left( X_{tau} \right)$, where $\tau \in \{\Delta,\dots, N-\Gamma\}$ . 

\section{The DNN architecture optimisation and performance analysis}\label{sec-analis}
\begin{figure}[h] 
\centering
    \includegraphics[width=\linewidth]{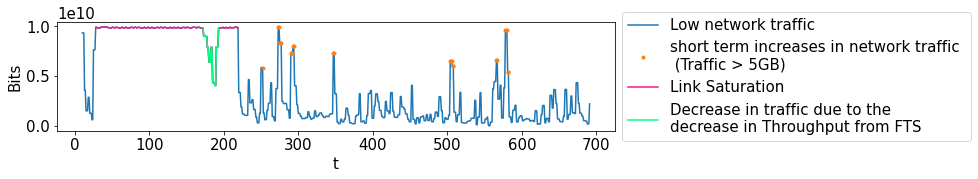}     
    \caption{Classification of the network traffic of the first data set into four types.} 
    \label{fig:traffic} 
\end{figure}

The DNN architecture hyper-parameters (including the number of filters in the convolutional layers, $f$) and batch sizes are optimised using the data presented in figure \ref{fig:testsub2}, which we reproduce in figure \ref{fig:traffic} to highlight its main characteristics. The pink region represents an overloaded link; a traffic drop, caused by a FTS throughput drop,  is visible in green. The juxtaposition of these two periods is called the $\Psi$ period. The orange points represent short term increases of traffic, above a general period of low network traffic, in blue. These last two features are neglected during optimisation.

We measure the DNN performance using a MSE (Mean Square Error), which is a popular choice as a loss function \cite{CNNLSTM} as  well as a common metric to evaluate the model accuracy \cite{gas}. We consider the one-step MSE and calculate an average MSE for the $\Gamma$-steps forecasting: 

\begin{equation}
   \text{MSE} = \text{MSE}\left(\Delta, \Gamma \right) = \sum^{\Gamma}_{i=0}\text{MSE}_{i},
\end{equation}
where $\text{MSE}_{i} = \frac{1}{N-\Gamma-\Delta}\sum^{N-\Gamma}_{t=\Delta}(y_{t,i} - \hat{y}_{t,i})^{2}$, for $i \in \{0, \dots , \Gamma \}$.

\begin{figure}[h]
\centering
\begin{minipage}{.35\textwidth}
  \centering
  \includegraphics[width=\linewidth]{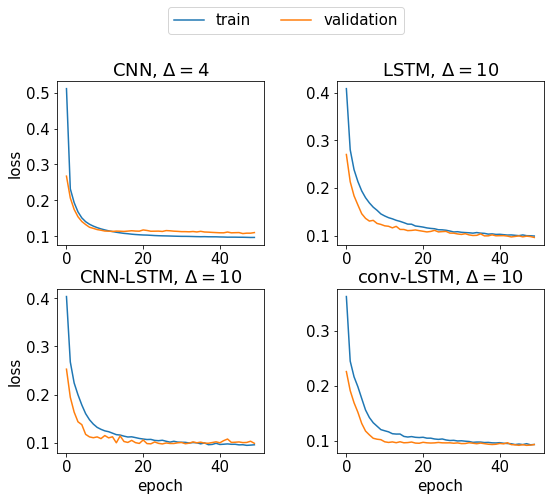}
  \captionof{figure}{Validation and training losses for the four models. Epoch = 50, b = 64, f = 8, ${\Gamma=5}$.}
  \label{fig:loss1}
\end{minipage}%
\hspace{0.5cm}
\begin{minipage}{.35\textwidth}
  \centering
  \includegraphics[width=\linewidth]{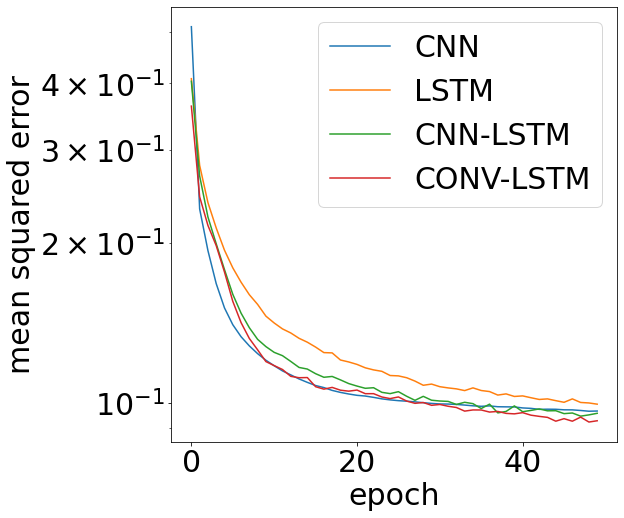}
  \captionof{figure}{Training losses. }
  \label{fig:loss2}
\end{minipage}
\end{figure}

\begin{figure}[h] 
\centering
    \includegraphics[width=\linewidth]{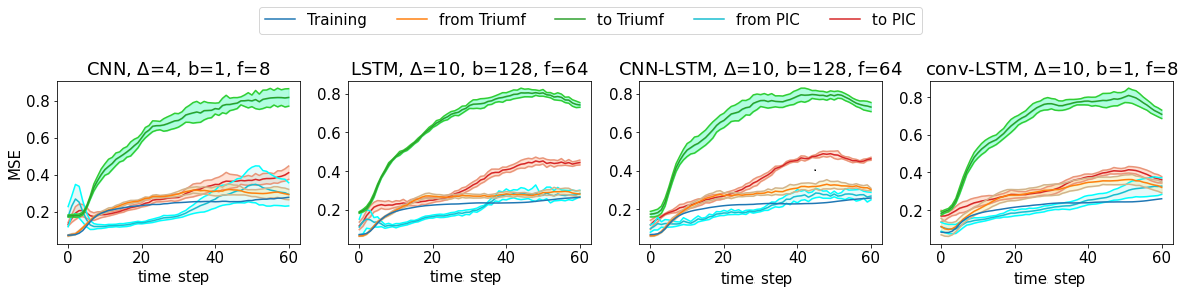}     
    \caption{Average MSE and its variance with respect to the forecasting steps (here: $\Gamma = 60$).} 
    \label{fig:diffrentdatset} 
\end{figure}

The $\Gamma$ value (i.e. the number of steps to predict) is  chosen according to the maximum error we can tolerate on the forecast (figure \ref{fig:diffrentdatset}). We chose a relatively small $\Gamma$ = 5 (10 minutes) taking into account that the validation data, like the  training data, do not contain a comparably long-term saturation. 
Figure \ref{fig:loss1} shows the loss functions for the different models and the best hyper-parameters:
 reducing the number of batches yields a better saturation prediction but seems to reduce sensitivity to small changes. The window size ($\Delta$) seems to have little effect on the LSTM's short-term predictions  while it significantly affects the CNN model.  However, it is worth remembering that, since the validation data do not include very long transfers, a larger window generates over-fitting: a behavior  visible, in particular, for the CNN model  using $\Delta$ = 4. For longer period prediction (at saturation) a larger window, corresponding to a longer input sequence, seems to perform better. In the convolutional models (CNN, Conv-LSTM) the dimensionality of the output space (filters) should be similar to the window value, and indeed we find that the best results are obtained for f = 8. The dimensionality of the output space (units) for the best result for LSTM and CNN-LSTM is f = 64.
In addition the loss function, presented in figure \ref{fig:loss2}, seems to suggest smaller prediction errors for models containing convolutions, which is in line with our expectations.
To check model stability, we measure the performance of 10 independent trainings. Figure \ref{fig:diffrentdatset} shows, for all data sets, the forecasting mean $MSE_{i}$ for $i\in\{0,\dots,\Gamma\}$ and its variance. In all cases, the low MSE variance indicates that model training is stable. For each time step, we obtained ~60 future values ($\Gamma = 60$). While the models perform similarly on the test data (from TRIUMF to CERN) as on the training data (collected along the same path and the same transfer direction), transfers to the TRIUMF site are not described correctly by any of the models. 

The accuracy of the network traffic  prediction appears to depend on the direction of transfers; transfers from TRIUMF, compared to transfers to TRIUMF, contain more non-FTS network traffic and so exhibit a larger MSE.
Transfers to and from PIC have a similar behaviour. It is worth noting, though, that the network traffic from PIC is easier to describe than that from TRIUMF, and it is predicted with a smaller MSE even during long-term forecasting. Perhaps this is because, compared to the training path, a greater fraction of the traffic between  PIC and CERN is generated by  FTS.
\vspace{-1cm}
\section{Results} \label{bestmodel}
\begin{figure}[h] 
\centering
    \includegraphics[width=0.9\linewidth]{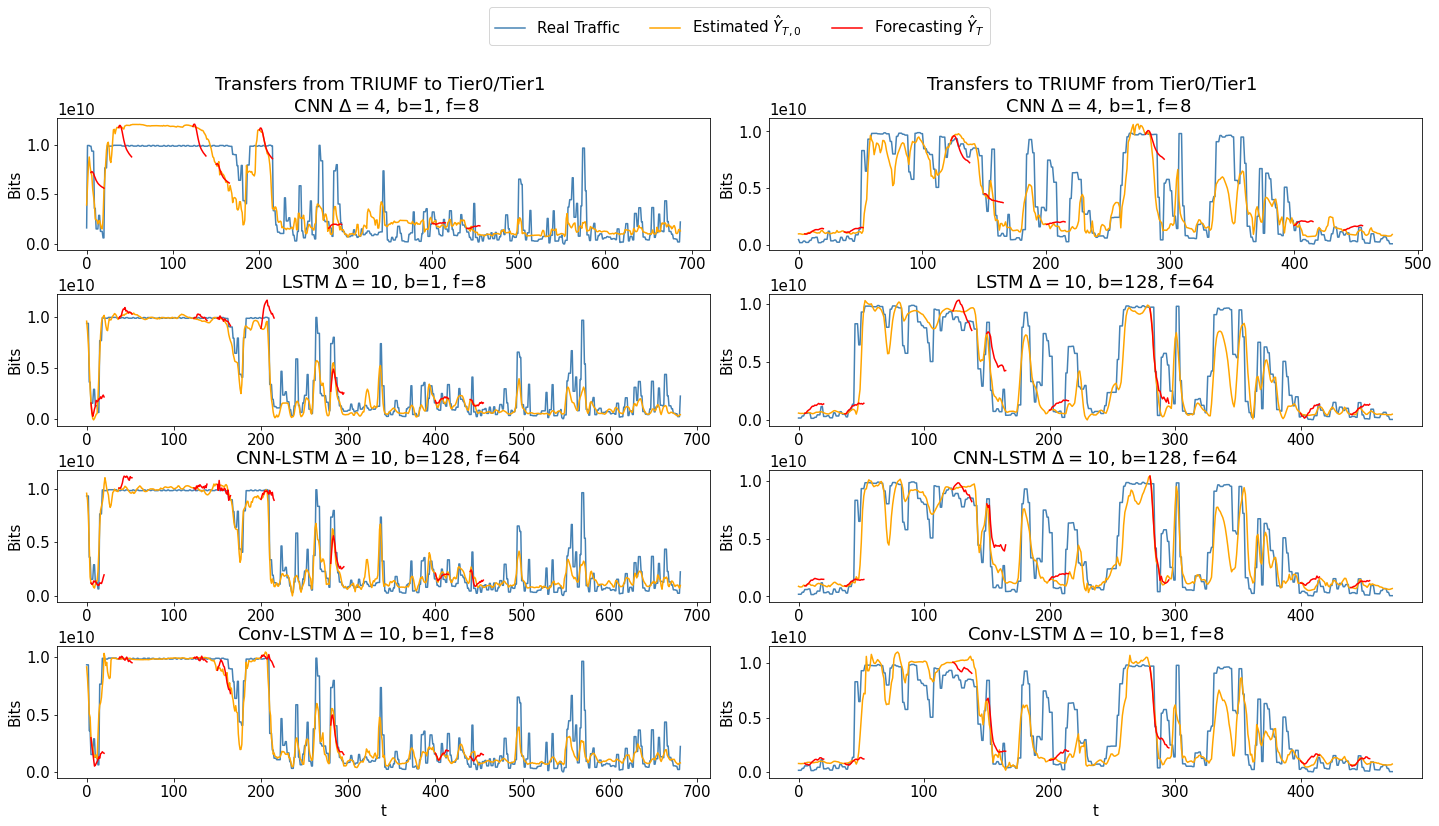}     
    \caption{Juxtaposition of two data sets and models.  $\hat{Y}_{\tau}$ is described by equation \ref{eqn:outputfortau} and is presented only for selected t. $\hat{Y}_{\tau,0}$ is 0th variable for $\hat{Y}_{\tau}$ for all t. } 
    \label{fig:triumftofrom} 
\end{figure}

\begin{figure}[h]
\centering
\hspace{0.5cm}
\begin{subfigure}{.35\textwidth}
  \centering
  \includegraphics[width=\linewidth]{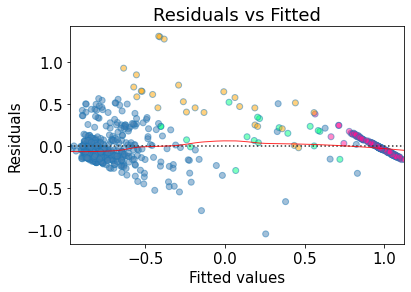}
  \caption{Fitted vs residuals}
  \label{fig:Residua2}
\end{subfigure}
\hspace{0.5cm}
\begin{subfigure}{.35\textwidth}
  \centering
  \includegraphics[width=\linewidth]{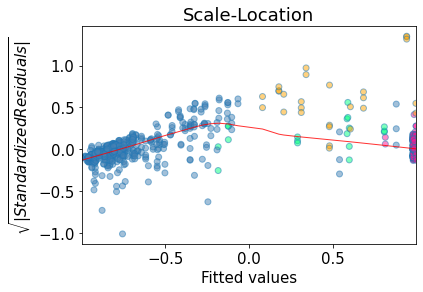}
  \caption{Real vs residuals}
  \label{fig:Residua3}
\end{subfigure}
\caption{Residuals - Effect of network traffic on the estimated (fitted) values (a) and on the real values (b)}
\label{fig:Residua}
\end{figure}

We chose the best model taking into account its performance on two different tasks. The first task describes the impact of transfers on network traffic at a given moment in time for $\Gamma \geq 0$, and it is evaluated by measuring the  $\text{MSE}_0$ (one-step of forecasting) in data set representing transfers from TRIUMF (equivalent to the train data). We chose the model that generates the smallest $MSE_0$. Since the $MSE_0$ was obtained for the whole data set (not only for the $\Psi$ period), it concerns both the estimation of long increased traffic (saturation) and the normal period. We expect that residuals may significantly depend on the traffic. Figure \ref{fig:Residua} confirms this hypothesis. The best models according to this definition are summarised in Table \ref{tab:traintestMSE} for ${\Gamma=15}$~(30~minutes): values are averaged over 10 training repetitions. The second task  takes into account the $\Psi$ period estimation (figure~\ref{fig:traffic}, table~\ref{tab:msels}) and the corresponding best model configurations are  shown in figure~\ref{fig:triumftofrom}. 

Considering the smallest $\text{MSE}_0$  in the test data, the CNN-LSTM with delta = 10 (CNN-LSTM(10)) seems to best describe the current network traffic. However, taking into account the prediction variance, the LSTM is also a good candidate. These models work the best for the entire data set, both on long-term and short term saturation. However the MSE on the  test data set is at a minimum in the  Conv-LSTM case. Note that the min(MSE$_0$) for the CNN-LSTM(10) model is smaller then the  min(MSE$_0$) for the  LSTM(10), but at the same time the MSE value, calculated on the test sample,  is bigger. This means that CNN-LSTM(10) can better explain the current traffic than LSTM(10), but LSTM can perform better on long term forecasting. This is because the time stamp ${\text{MSE}_i}$ increases more slowly for the LSTM(10) than for the CNN-LSTM(10) (figure \ref{fig:diffrentdatset}). On the other hand, the CNN-LSTM can better predict  minor irregularities: the model seems to be more sensitive to short-term increases in traffic caused by short transfers. Unfortunately the  CNN-LSTM also shows some over-fitting (probably due to the limited training data set size) which can explain why worse forecasting results are observed. The CNN-LSTM is also the  model providing  the most accurate estimate of the unexpected throughput drop,  in green in figure \ref{fig:traffic}.
The most interesting part of project is traffic estimation at the time when the link should be/is overloaded. Therefore, the results for the period $\Psi$ are presented in table \ref{tab:msels}, and the best models are selected based on this analysis.

\begin{table}[h]
\small
\caption{\label{tab:traintestMSE}Comparison of model parameters on the training set. $\Gamma  = 15$}
\centering
\begin{tabular}{|c|c|c|c|c|c|c|c|}
\hline
$\Delta$ & Model  &   \thead{ Batch -\\  Filters\textbackslash\\Units }   & MSE train & MSE test &  \thead{ $\text{MSE}_0$ \\ train} &  \thead{  $\text{MSE}_0$ \\test } & time [s]\\ \hline
4 & CNN   & 1-8 & 0.164   & 0.171 & 0.101 & 0.138   &    116     \\  \cline{2-8} 
& LSTM     & 128-64 &0.155   &0.155 &0.071 &  0.065 & 8  \\  \cline{2-8} 
& CNN-LSTM & 128-64 & 0.151  &0.154 &  0.069 &   0.068&  10  \\  \cline{2-8} 
& CONV-LSTM  & 1-8 & 0.167 & 0.161 & 0.086 &0.067   & 0.135    \\ \hline
10 & CNN      & 1-8 & 0.155   & 0.157 &  0.067 & 0.074  &  158    \\  \cline{2-8} 
& LSTM      & 128-64  & 0.151  & 0.150 & 0.062 &  0.056 &  19  \\  \cline{2-8} 
& CNN-LSTM   & 128-64 & 0.148 &  0.157&  0.058 & 0.054 &   19 \\  \cline{2-8} 
 & CONV-LSTM  & 1-8 &0.145  & 0.141 &  0.068 & 0.065 & 318 \\  \hline
\end{tabular}
\end{table}

\begin{table}[h]
\footnotesize
\caption{\label{tab:msels}Comparison of model parameters on the test data set representing transfers from TRIUMF to Tier0/Tier1. $\text{MSE}_{\Psi,0}$ and $\text{MSE}_{\Psi}$ means respectively $\text{MSE}_0$ and MSE average during $\Psi$ period. $\Gamma = 15$ (30 minutes). S is the standard deviation over 10 training repetitions.}
\centering
\begin{tabular}{|c|c|c|c|c|c|c|c|}
\hline
$\Delta$ & Model  & \thead{ Batch -\\  Filters\textbackslash\\ Units }   &  $\text{MSE}_{\Psi}(\Gamma)$ & $S(\text{MSE}_{\Psi}(\Gamma))$ &  $\text{MSE}_{\Psi,0}$& $S(\text{MSE}_{\Psi,0})$  &$r_s$ \\ \hline
4 & CNN   & 1 - 8  & 0.206 &  0.007 & 0.206 & 0.009&   0.764    \\  \cline{2-8} 
& LSTM    & 128 - 64 & 0.224 & 0.008  & 0.042 & 0.005 &   0.845 \\  \cline{2-8} 
& CNN-LSTM & 128 - 64& 0.233 & 0.015& 0.060 &  0.007 &   0.857   \\  \cline{2-8} 
& CONV-LSTM & 1 - 8 &  0.159 & 0.012  &0.048 &   0.007 & 0.864    \\  \hline
10 & CNN     & 1 - 8 & 0.223 &  0.095 &0.223 &0.010 &   0.787   \\  \cline{2-8} 
& LSTM      & 128 - 64 & 0.185 & 0.012 & 0.025 & 0.006&  0.841 \\  \cline{2-8} 
& CNN-LSTM   & 128 - 64 & 0.188 & 0.011 & 0.021 & 0.006 &  0.861  \\  \cline{2-8} 
 & CONV-LSTM &  1 - 8  & 0.125 &  0.008  & 0.036 & 0.008 &   0.862 \\  \hline
\end{tabular}
\end{table}

Although capable of estimating the network traffic trend, models containing only convolutional layers i.e: CNN(4) and CNN(10), perform worse on time series estimation. It is worth noting that such  models work better for a shorter term input  window. Differences between errors on the training data (both for the entire $Y$ prediction and only for the first $\Upsilon$ element) are large. In figure \ref{fig:triumftofrom}, we can see that the model has not learned how to predict long-term saturation of the link (each time it predicts a decrease with a similar rate of decrease). It also performs worse during general periods of low network traffic.

The LSTM  improves the accuracy of the one-step prediction task: all models correctly predict the link saturation caused by transfers from TRIUMF to Tier0 / Tier1. In addition, the LSTM  also performs very well during a period of increased network traffic. The CNN layer helps to detect small differences that improve the overall model sensitivity. This results in the best fit for the single-step forecast, while  it affects negatively the long-term forecast. 

As in the case of the entire data set (table \ref{tab:traintestMSE}), CNN-LSTM(10) exhibits the lowest MSE$_{\Psi,0}$. In order to investigate possible biases and distortions in the CNN-LSTM(10) prediction, we analyse the one-step forecasting residuals. Figure \ref{fig:Residua} shows that the that the data is heteroscedastic: the residuals distribution shows a large non-linear dependency between the volume of network traffic and the estimated data. Colour coding is as in figure \ref{fig:traffic}. Orange points are not  predicted with great accuracy, while, during the $\Psi$ period, residuals  accumulate around zero. The model rarely overestimates the traffic, while the data axhibit strong auto-correlation as shown by figure \ref{fig:Residua}.

Since residuals are not  linearly distributed, we use the the Spearman rank correlation~($r_s$) to measure the strength of an association between the two variables (real and estimated \mbox{traffic).} The Spearman coefficient is a non-parametric rank statistic, and assesses how well an arbitrary monotonic function describes a relationship between two variables, without making any assumptions about the frequency distribution of the variables. It is a special case of a Pearson correlation coefficient applied to the rank variables~\cite{hauke2011comparison}. In our test, using all models, the p-value is close to zero, thus the null hypothesis of uncorrelated samples is rejected. The measured Spearman rank coefficient (in table \ref{tab:msels}) suggests that the data belong to populations with different variances \cite{homebrew,HeteroskedasticityMat,YIN199069}. To confirm this hypothesis we  perform a Levene test on the  validation data: for all models, we  reject the null hypothesis (that all input samples belong to populations with equal variances) thus  confirming the heterogeneity property.

The main objective of our model is to detect the beginning and end of link saturation. Therefore, the MSE$_\Psi(\Gamma)$ and MSE$_{\Psi,0}$ should be considered. According to the results in table \ref{tab:msels}, the  Conv-LSTM model is preferred, since it provides a better estimate of the network traffic during the $\Psi$ period. MSE$_{\Psi}(\Gamma)$ is smaller and the variability of the estimated parameter is much lower. 

Furthermore, figure \ref{fig:triumftofrom} shows the performance of the model applied to a second, more difficult data set (figure \ref{fig:testsub2}): the Conv-LSTM model is not unduly affected around t=100. By ignoring  short-duration declines, we avoid  wrongly assuming that the transfer is ending and so reconfiguring the network back to its original status prematurely. The Conv-LSTM model is less sensitive to rapid changes, therefore, compared to CNN-LSTM, The Conv-LSTM is worse at detecting small transfers, which increases the MSE. 

\section{Conclusions and Future Work}\label{conlusion}

LSTM and CNN networks are among the most efficient and widely used deep learning models. Their use on time-series analyses is based on the LSTM ability to capture sequence pattern information and the CNN power in terms of data de-noising and feature extraction.  A CNN-LSTM combined model overcomes the CNN limitations  in terms of long temporal dependencies and achieves optimal forecasting performance.
Our work shows that CNN-LSTM and Conv-LSTM architectures can indeed enable us to detect network saturation and that they provide great forecasting accuracy even over long time periods (up to 30 minutes). 
 In addition, we provide a detailed performance comparison among different models, highlighting their strengths and flaws according to the specific task at hand.
 In particular, we are interested in accurately characterising large transfers, by predicting their occurrence and modelling their development, in order to be able to act in a manner that reduces link saturation.  Bearing this in mind, we consider CNN-LSTM as the best choice for transfer prediction and  Conv-LSTM  as the most suitable model to predict the transfer end. of relevance here is that Conv-LSTMs exhibit a slower reaction to drops, errors and repairable problems (problems that can automatically be adjusted by FTS) and, therefore, they can prevent multiple reconfiguration of the network during a short period of time.
 
 Future research should focus on further optimising the DNN architecture, in particular in terms of the relative strength between the CNN and LSTM components in the hybrid models. In addition, a systematic investigation of their capability to generalise to different data sets will not only, improve their usability for this specific application but also contribute to DNN interpretability and the understanding of the learning process. 

 
Previously, transfer detection did not always correctly characterise the evolution of network traffic. The use of models to characterise network traffic enables us to automate configuration actions to improve network utilisation efficiency and so reduce the duration of large data transfers.




%
%
%
\newpage
\bibliography{bibliography} 
%
%

\end{document}